\documentclass[nofootinbib, superscriptaddress, twocolumn]{revtex4}

\usepackage{amsmath}
\usepackage{amssymb}
\usepackage{color}
\usepackage{graphicx}
\usepackage{hyperref}
\usepackage[utf8]{inputenc}
\usepackage{listings}
\usepackage{url}

\newcommand{\code}[1]{\texttt{#1}}

\begin{document}

% Scalable
\title{Chemora: A PDE Solving Framework for Modern HPC Architectures}

\author{Erik Schnetter}
\homepage{www.perimeterinstitute.ca/personal/eschnetter}
\email{eschnetter@perimeterinstitute.ca}
\affiliation{Perimeter Institute for Theoretical Physics, Waterloo,
  ON, Canada}
\affiliation{Department of Physics, University of Guelph, Guelph, ON,
  Canada}
\affiliation{Center for Computation \& Technology, Louisiana State
  University, LA, USA}

\author{Marek Blazewicz}
\affiliation{Poznan Supercomputing and Networking Center, Poznan, Poland}

\author{Steven R. Brandt}
\affiliation{Center for Computation \& Technology, Louisiana State
  University, LA, USA}
\affiliation{Department of Computer Science, Louisiana State
  University, LA, USA}

\author{David M. Koppelman}
% \homepage{www.ece.lsu.edu/koppel}
% \email{koppel@ece.lsu.edu}
\affiliation{Center for Computation \& Technology, Louisiana State
  University, LA, USA}
\affiliation{Division of Electrical \& Computer Engineering, 
Louisiana State University, LA, USA}

\author{Frank Löffler}
\affiliation{Center for Computation \& Technology, Louisiana State
  University, LA, USA}

\date{2014-09-16}

\begin{abstract}
  Modern HPC architectures consist of heterogeneous multi-core,
  many-node systems with deep memory hierarchies. Modern applications
  employ ever more advanced discretisation methods to study
  multi-physics problems. Developing such applications that explore
  cutting-edge physics on cutting-edge HPC systems has become a
  complex task that requires significant HPC knowledge and experience.
  Unfortunately, this combined knowledge is currently out of reach for all
  but a few groups of application developers.
  
  Chemora is a framework for solving systems of Partial Differential
  Equations (PDEs) that targets modern HPC architectures. Chemora is
  based on Cactus, which sees prominent usage in the computational
  relativistic astrophysics community. In Chemora, PDEs are expressed
  either in a high-level \LaTeX-like language or in Mathematica.
  Discretisation stencils are defined separately from equations, and
  can include Finite Differences, Discontinuous Galerkin Finite
  Elements (DGFE), Adaptive Mesh Refinement (AMR), and multi-block
  systems.
  
  We use Chemora in the Einstein Toolkit to implement the Einstein
  Equations on CPUs and on accelerators, and study astrophysical
  systems such as black hole binaries, neutron stars, and
  core-collapse supernovae.
\end{abstract}

\maketitle

\section{Introduction}

Most computing hardware that is currently used for scientific applications
is highly parallel in nature. While this has been true for high
performance computing (HPC) systems for decades, this is now also true
for workstations, laptops, and even cell phones.

The Top500 list % \cite{top500:web}
-- contended as it is today
\cite{Dongarra:2013a} % hpcg:web
-- nicely demonstrates this. A modern
CPU, if programmed serially, may be able to execute instructions with
at most about 3~GFlop/sec. At the same time, a modern laptop already
has a theoretical peak performance of about 100 GFlop/sec, which is a
ratio of a factor of 30. Unsurprisingly, for the top three systems of
the current Top500 list, the performance ``gain'' from parallelism (as
opposed to serial single-core performance) is more than a factor of
$10^7$. This parallelism consists not only of multi-node and
multi-core architectures, but also of vector instructions (SIMD) and
superscalar instruction execution.

At the same time, mainstream languages used in scientific computing
(Fortran, C, C++, Maple, Mathematica, Matlab, Python, etc.) are
inherently serial. They may offer certain additions to incorporate
parallelism (numerical libraries, MPI, OpenMP, async, thread or
process pools, ParallelTable, etc.), but these additions are usually
coarse-grained and awkward to use.

Satish et al. \cite{Satish2012a} examine the performance obtained by
``naively written C/C++ code'', and compare this to the ``best
optimized code'' for a modern Intel processor. They call the
performance difference the \emph{Ninja Performance Gap}, and found
that this gap is \emph{a factor of 24 on average} --- and this is for a
single processor, {\it i.e.,} without taking MPI parallelism into account, or
without considering many-thread accelerators (GPUs). They go on to describe the
low-level code transformations necessary to close this gap. They
conclude, somewhat optimistically, that (future) hardware support and
advances in compiler technology will be able to close this gap. We do
not quite share this optimism.

Unfortunately, application developers are often unwilling to change
low-level details of their codes to adapt to different system
architectures, since this is a very time-consuming task. We also find
that current compiler technology is definitively not yet advanced enough
to automatically re-structure large, complex loop kernels
(implementing complex physics) to improve performance.
These
challenges and more are described in detail e.g. in \cite{ascr:2011}.

We structure this paper around three \emph{lessons learned}:
\begin{enumerate}
\item Designing and implementing large applications requires expertise
  in three disciplines: physics, mathematics (discretisation), and
  computer science (implementation). One needs some kind of software
  framework to structure respective collaborations among these domain
  experts.
\item Hardware and software are always changing. Current languages
  (Fortran, C, C++, OpenCL, CUDA) are too low-level to conveniently
  express the physics or mathematics in applications. One needs tools
  such as automated code generation to be able to quickly restructure
  fundamental elements of applications, such as loop kernels,
  differencing stencils, or the memory layout of data structures.
\item MPI and OpenMP become increasing difficult to use efficiently at
  large core counts, or in highly dynamic applications using adaptive
  mesh refinement (AMR) or unstructured grids. One needs more elegant
  ways to express fine-grained multi-threading, and to handle
  migrating data between nodes.
\end{enumerate}

In the following, we will describe our approach to addressing these
lessons. We also describe related work in each section. Our
software is available as open source.

\section{Cactus: Software Framework for High-Performance Computing}

Designing and implementing an application that simulates a non-trivial
physical system requires expertise not only in the respective
sub-field(s) of physics, but also expertise in numerical analysis to
properly treat the approximations made when discretising, and
expertise in computer science to arrive at an efficient application
that performs well on modern hardware and is parallelised to take
advantage of tens of thousands of nodes.

Enabling such collaborations is not a trivial task, since few research
groups are large enough to have all this expertise in house. 
Typically, parts of codes are designed and modified at different times,
during which collaborators may move to different
institutions. As a result, the collaborations are ad-hoc and often
informal. In addition, as graduate students continue their career and
become postdocs, and then maybe become faculty who lead their own
groups, previous collaborators become friendly competitors, and the
dividing line between collaboration and competition becomes fuzzy. Any
software framework for large-scale applications needs to address this
social aspect of high-performance computing.

Here we use the \emph{Cactus Software Framework} \cite{Goodale:2002a,
Cactuscode:web}, and we briefly describe how its design supports
large yet only loosely defined collaborations in practice. A more
detailed description can be found in \cite{Loeffler:2013aa};
references to other software frameworks and their approach to these
issues are listed e.g. in \cite{Dubey:2013a}.

Applications that use Cactus are written as a set of
modules/components/libraries/plugins (called \emph{thorns}) that are
connected by glue code in the framework (\emph{flesh}). The flesh
handles only metadata and does not touch the thorns' data structures,
ensuring that it does not get into the way of efficiency. The flesh
also provides certain basic services such as managing run-time
parameters, scheduling execution of routines contained in thorns, and
allowing introspection into other thorns' metadata to infrastructure
thorns (e.g. for checkpointing/recovery). Finally, the flesh contains
a \code{make}-based build system; anyone who has to build and install
a set of inter-dependent libraries on a modern HPC system will
appreciate the simplicity of building multiple thorns (libraries) with
a single command.

Thorns in a Cactus-based application implement not only the physics
and discretisation methods, they can also provide infrastructure
services such as domain decomposition, memory management, MPI
communication, I/O, or checkpoint/recovery. Externalising these tasks
into separate thorns significantly simplifies implementing physics
thorns. In practice, one can find two different kinds of routines in a
typical physics thorn: Algorithmic routines that make high-level
decisions (e.g. ``is another iteration necessary?''), and worker routines
that perform the heavy lifting, corresponding to a \emph{kernel} in
CUDA or OpenCL.

Cactus thorns can be written in different languages; currently
supported are C, C++, CUDA, Fortran, and OpenCL, with support for Lua
\cite{lua:language} % lua:web
under development.

\label{sec:driver}
The predominant parallelisation model of Cactus applications is today
based on a hybrid MPI+OpenMP scheme, where an MPI domain decomposition
is provided by a \emph{driver} thorn (and not implemented in physics
thorns), and where physics thorns use e.g. OpenMP or CUDA for
shared-memory parallelism (multi-threading). The driver thorn also
provides memory management, including transferring data from/to
accelerators.

Time stepping in Cactus is implemented by the driver as well (hence
its name). The driver first executes all initialization routines, and
then executes the time evolution routines in loop until the
termination criterion is satisfied (e.g. a certain simulation time is
reached). This could also be used to iteratively solve elliptic
equations. The actual time stepping algorithm, e.g. a Runge-Kutta
method, is implemented in a separate thorn that schedules its routines
to cycle/copy time levels, calculate the new state vector, or estimate
the time stepping error. The Cactus scheduler supports both
conditionals and loops, and \emph{named schedule groups} that
correspond to callback routines.

\subsection{Enabling Collaboration}

Thorns are combined into applications only by the end user, not by a
central authority. A Cactus release consists of a set of thorns that
can be combined in many ways, not of a single application. This gives
the end user complete flexibility over which components from which
developers to use. The connection points where thorns interact (e.g.
grid variables, schedule items, service functions) are explicitly
named, and these names and their meaning need to be standardised
within a community that intends to share thorns. This means that
thorns are \emph{self-assembling}, and it is not necessary to
explicitly describe the control flow or flow of information between
thorns. To create an application, one only lists the thorns it should
contain. Without self-assembly, combining thorns would be a laborious
task and this would defeat the purpose of such a design.

This is one of the \emph{key points} in the design of Cactus: It
allows every user to independently choose their collaborators, and
also allows them to implement new or modify existing thorns if certain
functionality is missing, without ever needing to incorporate their
work into ``the master branch''.

It turns out that, in practice, almost all Cactus users are also
Cactus developers implementing their own thorns, if only for new
initial conditions or analysis methods. Given that thorns are
connected only via the flesh instead of directly to each
other,\footnote{This paragraph describes the ideal design of Cactus
  components. Most components follow this design pattern. Exceptions
  are possible, and are sometimes necessary.} it is not necessary for
new thorns to be ``accepted'' by an authority or ``incorporated'' into
a new release. Many, if not all, research groups who use Cactus have
their own private software repositories, where they develop new physics
capabilities in complete secrecy, as is necessary to succeed in a
competitive academic environment. At the same time, many research
groups choose to collaborate on other thorns that are not related to
their core competencies. For example, an astrophysics group may choose
to participate in shared development of improved parallelisation or
I/O capabilities, while keeping the existence of a new radiation
transport module secret.

\subsection{Einstein Toolkit}

Based on the Cactus software framework, the computational relativistic
astrophysics community has designed and implemented the Einstein
Toolkit \cite{Loffler:2011ay, Moesta:2013dna, EinsteinToolkit:web}.
This is a collaboration of many of the leading research groups, with
currently more than 100 members from more than 50 institutions. The
Einstein Toolkit originated in 2010 from many existing, high-quality
modules, and development of certain additional modules was funded by
several collaborative NSF awards.

The glue that holds the Einstein Toolkit together is a set of core
modules that unambiguously define certain standards. These standards
were discussed and decided in a community process that started much
earlier, and they have been revised and refined several times. These
standards include details such as
\begin{itemize}
\item names and unambiguous definitions for certain \emph{physical
  quantities} such as the metric, curvature, mass density, velocity,
  etc.;
\item names and meanings of \emph{schedule points} while running
  simulations, such as for setting up initial conditions, choosing
  gauge conditions, evaluating the right hand sides (RHS) of the
  evolution equations, or calculating the hydrodynamical pressure;
\item definitions pertaining to important \emph{basic analysis steps}
  (e.g. finding horizons) that are needed by later analysis stages;
\item conventions for \emph{laying out data} onto the computational
  grid functions, defining e.g. how hydrodynamical fluxes are
  staggered.
\end{itemize}

While Cactus itself offers some basic standards, these only target
generic physics simulations. Many additional such standards needed to
be set that are specific to numerical relativity. Following these
standards makes additional thorns inter-operable. In some cases, code
developers opted to ignore some of these standards since they were too
limiting, and in some of those cases, the Einstein Toolkit standard
definitions were later revised to accommodate the additional
requirements. There is also a continuing effort to phase out parts of
the infrastructure that are outdated, i.e. that have been unused for
some time and where no future need is anticipated.

We consider the Einstein Toolkit to be a very successful endeavour,
cited in probably more than 200 publications and many student theses as
basis for the respective research.

\section{Efficient Collaborations vs. Efficient Code}

To allow efficient collaboration between different domain experts
(physicists, mathematicians, computer scientists), it is important
that they can implement their algorithms and methods into different
modules, and that they do not all have to work on the same few lines
in a loop kernel to make their respective contributions.

Unfortunately, the latter is just what happens in a straightforward
implementation. Take, as a simple example, the scalar wave equation in
first order form
\begin{eqnarray}
  \label{eq:wave}
  \partial_t u & = & \rho
  \\\nonumber
  \partial_t \rho & = & \delta^{ij} \partial_i v_j
  \\\nonumber
  \partial_t v_i & = & \partial_i \rho \quad \textrm{.}
\end{eqnarray}
Implemented via finite differencing, this leads to a loop such
as\footnote{To improve readability, we keep our examples overly
  simple, restricting ourselves to one dimension, second order
  accuracy, and a pseudo-C-like syntax. Real applications will be
  significantly more complex.}
\begin{lstlisting}
#pragma omp parallel for
for (i=1; i<N-1; ++i) {
  dt_u[i] = rho[i];
  dt_rho[i] = (v[i+1] - v[i-1]) / (2*dx);
  dt_v[i] = (rho[i+1] - rho[i-1]) / (2*dx);
}
\end{lstlisting}
These few lines of code express simultaneously the physics that is
simulated (the system of equations), the discrete approximation
(finite differencing), and the mapping onto hardware resources (memory
layout of the state vector, multi-threaded via OpenMP). In a real
application, there would also be explicit choices determined by
multi-node parallelisation (e.g. ghost zones for MPI communication),
or maybe explicit loop tiling to improve cache efficiency.

Obviously, mixing these different concerns that have very different
goals into the very same few lines of code makes it virtually
impossible to modify, improve, or re-design these aspects
simultaneously. For example, adding additional physics to a system of
equations requires the physicist to understand many details of how
discretisation and parallelisation are implemented. Changing from
second order to fourth order finite differencing, or from finite
differencing to finite elements, requires re-writing the entire loop
kernel, and likely large parts of the inter-node communication
routines. Switching from MPI+OpenMP to a different parallelisation
model (e.g. offloading to an accelerator) requires rewriting the
entire loop kernel in CUDA or with OpenACC.

Clearly, this is a large obstacle that impedes progress.
Correspondingly, many current large-scale applications have developed
some set of abstractions that partially ameliorate this, e.g. moving 
finite differencing stencils into functions or macros, or implementing
them via C++ template metaprogramming. However, this still falls short
of what is needed to grant sufficient independence to physicists and
computer scientists.

We choose to employ automated code generation to allow separation of
concerns. 

\subsection{Existing Code Generation Systems}

Here we give a brief overview over several code generation systems,
and compare them to our system \emph{Kranc} as described below in
section \ref{sec:kranc}.

The state of the art for automated code generation is especially
advanced for Continuous Finite Elements, maybe due to a very
elegant mathematical description in terms of Differential Forms. A set
of tools allows creating complete simulation codes with little more
input than the system of equations that should be solved. Different
from our approach, these tools usually employ unstructured meshes;
these allow much greater flexibility in discretizing the problem
domain, but come at a significant performance cost. Consequently,
efficient implementation of a stencil-based discretization is outside
their scope. Well-known examples for such tools are
\emph{FEniCS} \cite{Logg2012a}, % fenics:web
\emph{FreeFEM++} \cite{MR3043640}, % freefem:web
\emph{Liszt} \cite{DeVito2011a}, or
\emph{Sundance} (part of Trilinos)
\cite{Long2012a, Long2013a}. % trilinos:web

Other tools target stencil-based discretization methods, and are thus
much closer to Chemora.
\emph{Paraiso} \cite{Muranushi2012a} % paraiso:web
stands out as it is
implemented in the functional language Haskell. It is otherwise
similar in design to Chemora, and includes dynamic optimizations to
improve code performance. However, it lacks the high-level
transformations that we apply in Mathematica, as well as many of the
low-level stencil optimizations we apply when targeting GPUs.

There are many tools supporting stencil computation in which the user
enters a computation kernel while the tool manages iteration and the
delivery of data in a way suitable for the computation target. More
recent work has been targeted at GPU accelerators and most of these
systems perform execution-driven autotuning in which trial executions
are performed to find a good configuration \cite{lutz13, khan13,
  zhang12, holewinski12, Maruyama2011a, christen11, kamil10}.
\emph{PARTANS} autotunes for multi-GPU systems \cite{lutz13}, while
the work of Khan et al. \cite{khan13} considers variations in data
staging and also mixes heuristic and autotuning techniques reducing
some of autotuning's startup overhead. \emph{Patus} \cite{christen11}
allows the user to specify execution alternatives for the autotuner to
explore, the sort of programmer burdening that Chemora is designed to
avoid. For Chemora, the starting point is a differential equation
description, the user does not write stencil codes. Nevertheless
Chemora does generate stencil code and uses autotuning to find good
tile sizes. Chemora's autotuning is model driven, reducing startup
time.

In addition to such tools, there exist languages to describe either
equations or complete physics systems. Some of the tools listed above
define their own languages that are closely related either the
respective tool or discretization method.
However, we want to mention in particular
\emph{Modelica}
% \footnote{Modelica and the Modelica Association,
% \url{https://www.modelica.org/}.}
as a tool-indepent and discretization-independent way of describing
physics systems \cite{Elmqvist1978a}. % modelica:web
Modelica is very
similar in spirit to our language \emph{EDL} described below in
section \ref{sec:edl}. Modelica seems to be targeting ODEs rather than
PDEs, and lacks support for describing discretization methods except
for uniform grids. On the other hand, Modelica offers many features
that EDL lacks, such as units, type definitions, or composing models;
EDL regains some of these via the Cactus framework.

\subsection{Kranc: Automated Code Generation}
\label{sec:kranc}

Our code generation system is called \emph{Kranc} \cite{Husa:2004ip,
  Kranc:web}, and is based on Mathematica. Mathematica offers a
convenient high-level language, combining Lisp-like pattern-matching
facilities with a syntax that is easy to understand.

The basic workflow is as follows. A system of equations is described
in Mathematica, and is combined with a choice of discretisation. The
Kranc package expands this system to C++ (or CUDA, OpenCL, \ldots)
code, and performs certain performance-improving transformations along
the way. The generated C++ code is a complete, independent Cactus
thorn, and can be built and run in the usual manner.

In a collaboration, a physicist or a mathematician interested in the
system of equations or its discretisation would mostly work at the
level of a Mathematica script that calls Kranc. A computer scientist
interested in efficiency and performance would modify or add to some
of the transformation stages in Kranc, or would work on Kranc's code
templates that contain the OpenMP or CUDA specific code. In this
respect, Kranc is a full-scale compiler with a parser (Mathematica), a
middle-end that transforms code in several stages and applies
optimizations, and a code generator. Since Kranc generates output in a
high-level language (e.g. C++), it does not have to deal with very
low-level architecture details such as register allocation or
instruction selection.

\subsubsection{Physics System Description}

To describe a physics system, one needs not only to describe the
system of equations (the RHS of the PDEs), but also specify the state
vector, dependent quantities (e.g. pressure dependence on density and
temperature), constraint equations (if any), as well as run-time
parameters, and specify whether and which quantities to import from
other Cactus modules.

In Kranc, equations can be described in a high-level form using
abstract index notation (aka the ``Einstein summation convention''),
and one can declare tensor symmetries. Kranc distinguishes between
covariant and partial derivatives, and expands covariant derivatives
and Lie derivatives automatically.
This is described in detail in \cite{Kranc:web}.

Many methods for solving elliptic equations require evaluating the
Jacobian, i.e. calculating the derivatives of the equations with
respect to the state vector variables. Deriving the Jacobian from the
physics equations is a tedious task if performed manually. In
Mathematica, this can be implemented automatically in a
straightforward manner. Our our code generator does not provide
explicit support for this, as this is not needed to solve the
Einstein equations, but it is possible to incorporate arbitrary
Mathematica code.

\subsubsection{Discretisation Description}

Kranc currently supports Finite Differencing as its discretisation method.
Support for Discontinuous Galerkin Finite Elements (DGFE) methods is
available in a pre-production version. Other methods (e.g. higher
order Finite Volumes) could be added in a straightforward manner.
Particle systems are not supported by Cactus yet, but would also be
possible.

Arbitrary derivative operators can be defined, either in a stencil
notation that is expanded by Mathematica, or by providing macros or
functions to Kranc's run-time system that are then called. In the
stencil notation, e.g. the second derivative operator $[+1,-2,+1]/h^2$
is expressed as \code{(+1 shift\textasciicircum(-1) -2
  shift\textasciicircum0 +1
  shift\textasciicircum(+1))/dx\textasciicircum2}. ``Standard'' Finite
Differencing operators of arbitrary order are built-in.

Finite Differencing with arbitrary order of accuracy is available. The
order can either be determined when Kranc is run, or can be left as
run-time option which will then be handled efficiently.

\subsubsection{Code Transformations}

Since Kranc expects equations entered in Mathematica, one can use the
full range of Mathematica features when doing so. For example, when
setting up initial conditions or boundary conditions, it is
straightforward to use computer algebra to evaluate derivatives or
integrals, or to use Mathematica's numerical features to evaluate
approximations.

Kranc expands the user's input by expanding vectors and tensors into
their components, while respecting symmetries. For example, a second
partial derivative $\partial_i\partial_j\rho$ is entered as
\code{PD[rho,i,j]}, where Kranc knows that this expression is
symmetric in the indices \code{i} and \code{j}. A definition of the
form $v_i = \partial_i\rho$ is expressed as \code{v[i]->PD[rho,i]},
and is translated into three separate assignments for variables
\code{v1}, \code{v2}, and \code{v3}. Derivatives such as
\code{PD[rho,i]} are translated into macros or function calls.

Kranc removes unused intermediate variables, and can perform common
subexpression elimination (CSE) to try and reduce the number of
operations. Code is generated in terms of \emph{calculations}, which
correspond to loop kernels, or kernel functions in CUDA. With Kranc,
one can semi-automatically combine or split loop kernels (without
having to explicitly rewrite them), where Kranc ensures that the
resulting kernels remain correct; it automatically removes unnecessary
terms, or duplicates them as necessary if kernels are split.
This is an important optimisation when a system of equations is too
large to fit into a CPU's cache because it either uses too much data
or contains too many instructions.\footnote{When evolving the Einstein
  equations, this is in fact the most important performance
  optimization.}

Finally, Kranc can explicitly vectorize a calculation by translating
all mathematical operations such as \code{+} or \code{*} into
CPU-specific intrinsics. In the end, certain peephole optimisations
are applied, e.g. eliminating double negations ($(-a)*(-b)$ becomes
$a*b$), combining multiplications and additions into a single
multiply-add operation ($a*b+c$ becomes $\textrm{mad}(a,b,c)$), or
replacing divisions by multiplications ($a/b/c$ becomes $a/(b*c)$).
While one may be hoping that compilers would these days perform these
simple optimisations, the reality is that many compilers do
not,\footnote{We regularly check the generated machine code.} and
Mathematica's pattern matching facilities make these
micro-optimisations very easy to implement.

After these transformations, the code is output as C++ with OpenMP,
CUDA, or OpenCL. The syntax of these languages is so similar that one
needs to make only minor changes during code generations. We used to
support Fortran as well, but found that (a) there was no measurable
difference in speed for Fortran and C without low-level optimisations,
and (b) many of these lower level optimisations could not be applied when
generating Fortran code.

We want to stress that implementing Kranc in Mathematica, as opposed
to using C macros or C++ template metaprogramming, makes it
significantly easier to add additional transformations or
optimizations to Kranc. Mathematica's Lisp-like pattern matching
functionality is ideally suited for this, and the respective
transformation rules are easily understandable also for
non-computer-scientists.

In addition to the transformation applied by Kranc, there exists a
non-trivial run-time library to efficiently map loop kernels to GPU
hardware. The library performs run-time model-driven auto-tuning to
optimize thread assignment based on parameter values and performs
dynamic compilation to minimize code and register overhead. Many CUDA
specific optimisations are implemented there, such
as to use the fast local memory of Nvidia GPUs efficiently. These
optimisations are described in \cite{ES-Blazewicz2012a} and
\cite{ES-Blazewicz2013a}, and their implementation is available at our
web site \url{chemoracode:web}.

\subsection{Equation Description Language}
\label{sec:edl}

While describing systems of equations and their discretisation in
Mathematica works very well in practice and has many advantages, there
are also drawbacks. Among those are:
\begin{itemize}
\item One has all the power of Mathematica, which makes it easy for
  beginners to make a mistake that is difficult (for them) to spot.
\item The input to Kranc is essentially a single, large data
  structure, describing variables, parameters, equations, etc.
  Mathematica's loose type checking rules mean that errors in setting
  up such a data structure are not always obvious, and if so, cannot
  be attributed to a specific line and column number.
\end{itemize}
To address this, we have designed an Equation Description Language
(EDL). This is a simple, \LaTeX-like language that is easy to parse,
and can readily be translated e.g. into an input for Kranc, or also
for other code generation systems. Since the EDL is read by a true
parser \cite{Brandt2010a, piraha:web}, errors lead to understandable
error messages with a line and column number.

\begin{figure}
  \begin{lstlisting}
# Evolved variables (state vector)
begin group Evolved
  u  : "scalar"
  rho: "rho-dot"
  v_i: "grad rho"
end group

# Extra variables (analysis quantities)
begin group Extra
  eps: "energy density"
end group

# Run-time parameters
begin parameters
  A: real "initial amplitude"
  W: real "initial width"
end parameters

# Calculations
begin calculation Init
  u   = 0
  rho = A exp(-1/2 (r/W)**2)
  v_i = 0
end calculation Init
begin calculation RHS
  D_t u   = rho
  D_t rho = delta^ij D_i v_j
  D_t v_i = D_i rho
end calculation
begin calculation Energy
  eps = 1/2 (rho**2 + delta^ij v_i v_j)
end calculation

# Discretisation
begin derivatives
  D_i = FiniteDifferencingOperator[1,1,i]
end derivatives
  \end{lstlisting}
  \caption{The scalar wave equation, expressed in our EDL. This
    corresponds to the formulation described in eq. (\ref{eq:wave})
    above. This is the complete input necessary to generate a complete
    Cactus thorn. Note that it contains the formulation of the system
    of PDEs, as well as a description of the discretisation method,
    here centered second-order accurate Finite Differencing. This
    description is easy to understand. (The design of the EDL is not yet
    finalised, and detail of the language syntax may change in the
    future.)\\
    \hspace*{1em} Note that, this description is free of details
    regarding the memory layout of data structures, the order in which
    loops are traversed, cache optimizations, or parallelisation;
    these choices are made elsewhere.}
  \label{fig:wave-edl}
\end{figure}

Figure \ref{fig:wave-edl} shows how eq. (\ref{eq:wave}) above reads in
our EDL.

\subsection{Single-Node Performance}

The methods described in this section -- specifying equations and
discretisation at a high level, and employing automated code
generation -- are still independent of distributed-memory parallelism
(i.e. MPI). However, they are important to achieving a high
single-node performance while still retaining the flexibility to
modify the system of equations.

This approach was crucial for the Einstein Toolkit to achieve good
single-node performance for the Einstein equations. Before employing
automated code generation, we used a hand-written Fortran code that
implemented finite differencing operators via macros
\cite{Alcubierre:2004hr}. This code typically achieved less than 5\%
of the theoretical peak performance.

Given the complexity of the Einstein equations (several thousand
floating point operations to evaluate the RHS at a single grid point),
experimenting with low-level transformations to improve performance
was deemed too tedious. Similarly, translating the code manually to
CUDA or OpenCL was never attempted.

After switching to automated code generation \cite{Brown:2008sb}, the
first versions of the generated code were only about half as fast as
the previous Fortran version. This performance difference turned out
to be caused by incidental (and accidental) design decisions. After a
few iterations of improvements to the code generator, the
auto-generated loop kernels now run at almost 20\% of the theoretical
peak performance under ideal (i.e. benchmarking) conditions. The two
most important optimisations were loop fission to not overflow the
instruction cache, and manual vectorisation. We note that the full
application has additional costs such as inter-process communication
or mesh refinement operations that are not counted in these numbers.

\section{Fine-Grained Multi-Threading}
\label{sec:pugh-bgp}

Supercomputers today rely on distributed-memory parallelism. The
standard programming model for such systems is that of
\emph{communication sequential processes}, and the standard
implementation tool is the Message Passing Interface (MPI). MPI is
widely used not because it is easy to use, but because it has been
shown that MPI makes it possible to achieve very good performance,
if one invests sufficient effort.

Cactus was designed with MPI in mind. In principle, parallelism in
Cactus is externalised to a driver (see section \ref{sec:driver}
above), but in practice the Cactus API was designed for communicating
sequential processes.

Cactus' original driver \emph{PUGH} supports only uniform grids, i.e.
neither mesh refinement nor multi-block methods. PUGH shows excellent
scalability to more than 100k MPI processes \cite{cactusbgp:web}.

However, most physics applications using Cactus today employ more
sophisticated discretisations than a uniform grid. \emph{Carpet}, a
more modern Cactus driver \cite{Schnetter:2003rb, Schnetter:2006pg},
% CarpetCode:web
supports both mesh refinement and multi-block
methods. Carpet is being used for simulations with 10k MPI processes,
or about 100k cores when using the hybrid MPI+OpenMP model
\cite{ES-Schnetter2013a}. Unfortunately, simulations employing
adaptive mesh refinement with Berger-Oliger style sub-cycling in time
are in our infrastructure currently limited to using about 1k MPI
processes (or 10k cores), as the serial processing of different
refinement levels inherent in Berger-Oliger AMR severely limits the
available amount of parallelism.

We are currently developing a new driver for Cactus that is based on
fine-grained multi-threading, and which should improve the scalability
of Cactus-based applications; see e.g. \cite{Luitjens2008a,
  VanStraalen2009a} for similar projects where such an approach was
successful.

\subsection{Improving Communication Performance}

If one employs a very simple performance model for inter-node
communication, then communication speed is limited either by
\emph{bandwidth} or by \emph{latency}. If a code is bandwidth limited,
then one is transferring too much data. There is often not much one
can do to remedy this via software engineering; instead, one needs to
switch to a different algorithm that requires less data to be
transferred.

If a code is latency limited, however, then there may be a solution:
One can run many tasks or threads within each node, so that other
tasks or threads can execute while some are waiting on communication.
This requires software parallelism (task/thread counts) \emph{much
  larger} than the available hardware parallelism (core count) to hide
the communication latencies.

We want to make a distinction here between two similar concepts,
namely task-based parallelism and fine-grained multi-threading. Both
describe ways to parallelise a code, and both apply within a single
node. Both would be combined with using MPI (or an equivalent
mechanism) for inter-node communication.
\begin{itemize}
\item We define \emph{task-based parallelism} as a design where each
  task has a well-defined dependency on results from other tasks (or
  on values received from another node). \emph{Once started, a task
    runs to completion} without interruption. Tasks may start other
  tasks. There may be an explicit schedule that describes the order in
  which tasks are run.
  
  This model is implemented e.g. in OpenMP's \code{parallel for}
  directive, in CUDA, OpenCL, or also in Charm++ \cite{Kale1996a,
    Acun2014a}, % charm:web
  Legion \cite{Bauer2012a}, % legion:web
  or
  Uintah \cite{Luitjens2008a, Berzins2012a, Berzins2013a}. % uintah:web
\item We define \emph{fine-grained multi-threading} as a design where
  threads do not need to have pre-defined dependencies.
  Threads may wait on results from other
  threads at any time, and \emph{are suspended while they are
    waiting}. Thread scheduling is only decided dynamically.
  
  This model is implemented e.g. in the pthreads API or in HPX
  \cite{Kaiser2009a, Tabbal2011a} % hpx:web
  \footnote{It it is worthy of
    note that the HPX API allows one to write both \emph{fine-grained}
    code and \emph{task-based} code, and to transform the former into
    the latter in many cases.} % \cite{hpx:talk13}
\end{itemize}

A multi-threading system fundamentally needs to have the capability to
suspend a running thread, and run other threads while a thread is
waiting. This is a significant hurdle to its implementation, making
task-based systems much easier to implement. At the same time, true
multi-threading systems are much easier to use in an application since
one does not have to decide on the threads' dependencies ahead of
time. In certain cases, this allows code to be written in a more
natural style.

Our current design ideas revolve around the same concepts as those
present in HPX. This extends C++11's multi-threading
facilities (\code{async}, \code{future}) and memory management
facilities (\code{shared\_ptr}) to distributed memory systems.
Improving distributed memory scalability is somewhat orthogonal to
achieving good single-node performance, and we have so far reached
production quality only for the latter.

\section{Application: Core-Collapse Supernova Simulations}

\begin{figure}
  \includegraphics[width=0.48\textwidth]{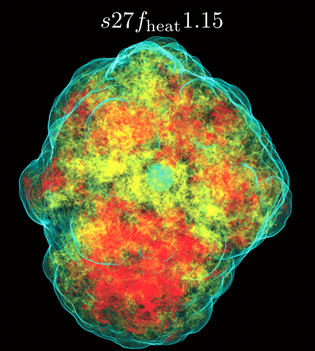}
  \caption{3D volume rendering of a simulated core-collapse supernova.
    This figure shows the specific entropy 150~ms after core bounce.
    Note the large scale global asymmetries and the many small
    blob-like protrusions in the shock front, which indicate that
    three-dimensional simulations are necessary to understand these
    systems. Image taken from \cite{Ott:2012mr, 1210.6674:web}; for
    physics details see there.}
  \label{fig:core-collapse}
\end{figure}

The science applications driving development of Chemora include the
study of black hole binaries, neutron stars, and core-collapse
supernovae. In these systems, gravitational effects are described by
general relativity, i.e. one needs to solve the Einstein equations.
These are a complex system of coupled, non-linear PDEs. In addition to
the Einstein equations, one also needs to solve the
general-relativistic hydrodynamics or magneto-hydrodynamics equations.
Figure \ref{fig:core-collapse} shows a 3D volume rendering of a
snapshot of a core-collapse supernova simulation, taken from results
published in \cite{Ott:2012mr, 1210.6674:web}.

% \begin{figure}
%   \includegraphics[width=0.48\textwidth]{Entropy_volren_0123_small}
%   \caption{
%     \cite{Mosta:2014jaa, 1403.1230:web}
%   }
% \end{figure}

In the code used for this simulation, the Einstein equations are
implemented via the Chemora framework, while the hydrodynamics
equations are still implemented manually. The high-level source code
describing the Einstein equations, important analysis quantities, and
their discretisation is about 1,500 lines long. The generated Cactus
thorn contains more than 40,000 lines of code.

It goes without saying that these simulations require significant
computing resources. Still, they are currently unable to include
important physical effects -- in particular, neutrino radiation
transport models will be needed to model core-collapse supernova
explosions in a self-consistent manner, and will increase the overall
computational cost by roughly an order of magnitude
\cite{ES-Schnetter2007a, ES-Ott2009a, ES-Ott2012b}.

\begin{figure}
  \includegraphics[width=0.48\textwidth]{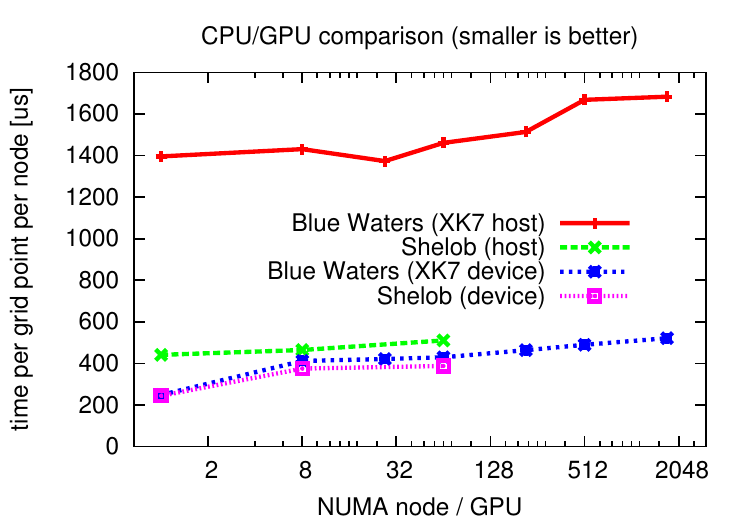}
  \caption{Performance comparison for a weak scaling experiment;
    smaller values are better. This compares two machines (Blue Waters
    at the NCSA, and Shelob at LSU), each with Nvidia GPUs. Our
    benchmark evolves the Einstein equations, a complex system of PDEs
    that is difficult to implement on GPUs (details see text). The $x$
    axis counts NUMA nodes (``sockets''/``processors'', or GPUs), the
    $y$ axis marks the amortized CPU time to evaluate the Einstein
    equations for a single grid point. On Blue Waters, our application
    scales to more than 2k nodes (32k cores), and achieves
    reasonable performance on the GPUs.}
  \label{fig:scaling}
\end{figure}

Finally, we show in figure \ref{fig:scaling} a performance comparison
for evolving the Einstein equations on typical CPUs and GPUs. This
weak scaling test uses a uniform grid without mesh refinement. In this
case, Cactus scales well up to at least 32k cores on Blue Waters. The
differences in run time between Blue Waters and Shelob are caused by
the differences in the respective CPUs' theoretical peak performance,
and by the fact that we are unfortunately not yet obtaining a good
floating point efficiency on Blue Waters' new AMD CPU architecture.
% Note that the Cactus infrastructure has shown to scale to more than
% 100k cores (see section \ref{sec:pugh-bgp} above).

\section{Summary}

Over the past decade plus of work, the developers of the Cactus
framework faced a series of challenges that we describe here not as
technology challenges, but as sociological challenges to enable
informal collaborations between researchers from different disciplines
such as physics, mathematics, and computer science.

Ensuring that scientific codes remain flexible is difficult, yet it is
necessary to ensure they will remain interesting to researchers with a
wide range of backgrounds. This in turn allows growing large-scale
applications requiring expertise in physics (equations), mathematics
(discretisation), and computer science (efficient implementation). One
needs some kind of framework to structure these collaborations, and
this framework needs at the same time to stay out of the way of
impeding code efficiency.

As hardware and software are changing, it is becoming clear that
current languages (C, C++, Fortran, OpenCL, CUDA, \ldots) are too
low-level to express modern ideas in physics systems, discretisation
methods, or how to map algorithms to hardware. We present automated
code generation as a simple-to-use and simple-to-understand mechanism
to be able to quickly restructure loop kernels, both to modify the
physics or discretisation, or to adapt it to new hardware.

Finally, it is widely accepted that MPI has become increasingly
difficult to use efficiently for dynamic applications on large core
counts. One needs a different abstraction layer that may or may not be
built on top of MPI. We envision a distributed-memory generalisation
of fine-grained multi-threading as solution suitable for the Einstein
Toolkit.

\begin{acknowledgments}
  We thank Peter Diener, Roland Haas,
  Ian Hinder, and Jian Tao as the main
  authors of Chemora. We also thank Barry Schneider for organising the
  session \emph{New Computational Techniques for Astrophysics} at the
  2014 APS April meeting where this paper was presented.
  
  Cactus, the Einstein Toolkit, and Chemora are currently supported by
  NSF awards 0905046 \emph{PetaCactus: Unraveling the Supernova --
    Gamma-Ray Burst Mystery}; 0941653 \emph{Enabling Science at the
    Petascale: From Binary Systems and Stellar Core Collapse To
    Gamma-Ray Bursts}; 1212401, 1212426, 1212433, 1212460 \emph{The
    Einstein Toolkit -- An Open-Source General Relativistic
    Multi-Physics Infrastructure for Relativistic Astrophysics};
  1265449, 1265434, 1265451 \emph{Using PDE Descriptions To Generate
    Code Precisely Tailored To Energy-Constrained Systems Including
    Large GPU Accelerated Clusters}; as well as NSERC award RGPIN
  418680-2012.

  Our research was made possible by access to HPC resources at ALCF,
  Blue Waters (NCSA), LONI, NERSC, NICS, SHARCNET, TACC, and via XSEDE
  allocations, as well as HPC systems at the Albert Einstein
  Institute, Caltech, Louisiana State University, and the Perimeter
  Institute.
\end{acknowledgments}

\bibliographystyle{apsrev-titles}
\bibliography{publications-schnetter,einsteintoolkit,cise}

\end{document}